NIH-PA Author Manuscript

# Integration of the OpenIGTLink Network Protocol for Image-Guided Therapy with the Medical Platform MeVisLab

**Jan Egger**,
Brigham and Women's Hospital and Harvard Medical School, Department of Radiology;
University of Marburg, Math. and Computer Science; University of Marburg, Neurosurgery

**Junichi Tokuda**,
Brigham and Women's Hospital and Harvard Medical School, Department of Radiology

**Laurent Chauvin**,
Brigham and Women's Hospital, Radiology

**Bernd Freisleben**,
University of Marburg, Math. and Computer Science

**Christopher Nimsky**,
University of Marburg, Neurosurgery

**Tina Kapur**, and
Brigham and Womenï¿½s Hospital, Surgical Planning Laboratory, Department of Radiology

**William Wells**
Brigham and Womenï¿½s Hospital, Surgical Planning Laboratory, Department of Radiology

## Abstract

We present the integration of the OpenIGTLink network protocol for image-guided therapy (IGT) with the medical prototyping platform MeVisLab. OpenIGTLink is a new, open, simple and extensible network communication protocol for IGT. The protocol provides a standardized mechanism to connect hardware and software by the transfer of coordinate transforms, images, and status messages. MeVisLab is a framework for the development of image processing algorithms and visualization and interaction methods, with a focus on medical imaging. The integration of OpenIGTLink into MeVisLab has been realized by developing a software module using the C++ programming language. As a result, researchers using MeVisLab can interface their software to hardware devices that already support the OpenIGTLink protocol, such as the NDI Aurora magnetic tracking system. In addition, the OpenIGTLink module can also be used to communicate directly with Slicer, a free, open source software package for visualization and image analysis. The integration has been tested with tracker clients available online and a real tracking system.

**Background—**OpenIGTLink is a new, open, simple and extensible network communication protocol for image-guided therapy (IGT). The protocol provides a standardized mechanism to connect hardware and software by the transfer of coordinate transforms, images, and status messages. MeVisLab is a framework for the development of image processing algorithms and visualization and interaction methods, with a focus on medical imaging.

**Conflict of interest statement** All authors in this paper have no potential conflict of interests.

**Methods**—We present the integration of the OpenIGTLink network protocol for IGT with the medical prototyping platform MeVisLab. The integration of OpenIGTLink into MeVisLab has been realized by developing a software module using the C++ programming language.

**Results**—The presented integration was evaluated with tracker clients that are available online. Furthermore, the integration was used to connect MeVisLab to Slicer and a NDI tracking system over the network. We also measured the latency time during navigation with a real instrument to show that the integration can be used clinically.

**Conclusions**—Researchers using MeVisLab can interface their software to hardware devices that already support the OpenIGTLink protocol, such as the NDI Aurora magnetic tracking system. In addition, the OpenIGTLink module can also be used to communicate directly with Slicer, a free, open source software package for visualization and image analysis.

## Introduction

Image-guided therapy (IGT) represents the use of medical images obtained either during or prior to a treatment, based on the assumption that knowledge of the location and orientation of a therapeutic process will allow a more specific therapy [1]. In the past few years, image-guided therapy has been successfully applied to many different clinical applications including aneurysm surgery [2], deep brain stimulation (DBS) [3], [4], robotic radiation delivery for the treatment of liver metastases [5], biopsy [6], [7], and radiotherapy (IGRT) of prostate cancer [8]. Locating surgical tools relative to the patient's body by using position and orientation tracking systems is now common. This can be achieved with optical [9], electromagnetic [10] or ultrasonic [11], [12] sensors. Furthermore, this can also be achieved with image acquisition using real-time ultrasound, computed tomography (CT) or magnetic resonance imaging (MRI). For visualization and guidance, this localization and image information is transferred from acquisition devices to navigation software. However, among the devices and the software in the operating room (OR) environment, standardization of the communication is still a common issue in image-guided therapy [13]. Furthermore, with the increasing number of medical software applications that are developed by researchers worldwide under different medical (prototyping) environments, such as Slicer [14], MeVislab [15], OsiriX [16], the XIP-Builder [17] – and the previous version of the XIP-Builder, the RadBuilder [18], [19] – standardization of information and communication technology is of increasing importance in order for these research applications to robustly interact with the devices in the operating room. Such standardization will also enable communication and information exchange between applications that have been developed for different medical environments; it will allow different research teams to work together to collaboratively solve open image-guided therapy problems while using their individual software development platforms. Furthermore, there has been a strong demand for communication standards among devices and navigation software with increasing research on robotic devices that support image-guided interventions to allow sharing of information such as target positions, images and device status. In order to tackle this problem, Tokuda et al. [20] defined an open, simple and extensible peer-to-peer network protocol for IGT called OpenIGTLink.

The advantage of OpenIGTLink is its simple specification, initially developed through a collaboration of academic, clinical and industrial partners for developing an integrated robotic system for MRI-guided prostate interventions [21]. It was designed for use in the application layer on the TCP/IP stack, allowing researchers to develop a prototype system that integrates multiple medical devices using the standard network infrastructure. Unlike existing interconnection standards for medical devices e.g. ISO 11 073/IEEE 1073 Standard for Medical Device Communication [22], CANOpen (EN 50 325-4) [23], Controller-Area Network (CAN) (ISO 118 988) [24], and Medical Device Plug-and-Play (MD PnP) [25], the OpenIGTLink protocol itself does not include mechanisms to establish and manage a session. It only defines a set of messages, which is the minimum data unit of this protocol. An OpenIGTLink message contains all information necessary for interpretation by the receiver. The message begins with a 58-byte header section, which is common to all types of data, followed by a body section. The format of the body section varies by the data type that is specified in the header section. Since any compatible receiver can interpret the header section, which contains the size and the data type of the body, every receiver can gracefully handle any message, even those with an unknown data type, by ignoring incompatible messages without the system crashing. Hence, this two-section structure allows developers to define their own data types while maintaining compatibility with other software that cannot interpret their user-defined data types. This simple message mechanism eases the development of OpenIGTLink interfaces and improves compatibility, thus it is suitable for prototyping a clinical system consisting of multiple devices and software connected via standard TCP/IP network. For a detailed description of the standard data types, see the publication of Tokuda et al. [20]. Further information is available on the web page provided by the National Alliance for Medical Image Computing (NA-MIC) [26].

In this paper, we present the integration of OpenIGTLink with the medical prototyping platform MeVisLab. MeVisLab is a framework for the development of image processing algorithms and visualization and interaction methods, with a focus on medical imaging. The integration of OpenIGTLink into MeVisLab has been realized by developing a software module in the C++ programming language. As a result, researchers using MeVisLab now have the possibility to connect to hardware devices that already support the OpenIGTLink protocol, such as the NDI Aurora magnetic tracking system. In addition, the OpenIGTLink module can also be used to communicate directly with Slicer, a free, open source software package for visualization and image analysis. The integration has been tested with tracker clients available online. Moreover, the integration was used to connect MeVisLab to Slicer and a commercial NDI tracking system over the network.

## Material and methods

This section describes our approach for integrating OpenIGTLink with MeVisLab. Thereby, the integration has been realized as a client-server model. A client-server model is a distributed application where a resource or service is provided by a server to a resource or service requester which is called the client. The basic concept of OpenIGTLink and the interaction of OpenIGTLink with different devices is shown in Figure 1: Image-guided therapy often relies on communication among sensors, devices and computers. For example, an intraprocedural imaging scanner e.g. MRI (see upper right image in Figure 1) transfers real-time images to navigation software (see upper left image in Figure 1) to allow the clinicians to monitor the progress of procedure; positions of surgical instruments are tracked by a optical position sensor device (see lower left image in Figure 1) and transferred to navigation software to indicate where those instruments are with respect to pre-procedural images; a robotic interventional assistance device (see lower right image in Figure 1) receives commands from navigation software and returns the current position of its end-effector or the current status of the device as feedbacks. Keeping Figure 1 in mind, the overall workflow of the OpenIGTLink / MeVisLab integration starts with the image data (see Figure 2). The image data is provided to a user-defined MeVisLab network. To load image data into a user-defined MeVisLab network, several modules such as OpenImage or ImageLoad exist that allow us to process various image formats, including the standard format for medical image data, called DICOM (Digital Imaging and Communications in Medicine, http://medical.nema.org). Besides the data fields (for example information about the images and diagnostic findings) DICOM also defines the syntax and semantic of commands and messages between DICOM compatible hardware. Nowadays, DICOM is the most used format for medical image data for communication between commercial software, open-source software, and devices/imaging systems. To continue processing the loaded image data, MeVisLab offers a collection of modules, such as image processing modules, visualization modules and also ITK (Insight Toolkit) (http://www.itk.org) modules [27]. In addition, MeVisLab allows users to integrate their own modules, for example, under Microsoft Visual Studio in C++. For the integration of the OpenIGTLink protocol with MeVisLab, a new module has been developed. As shown in Figure 2, the OpenIGTLink module handles the data exchange between MeVisLab and an external device such as a robot controller or 6-DOF position and orientation tracker. During the data exchange, different types of information such as the status, the image data or the position coordinates are transferred via the OpenIGTLink protocol. The overall workflow of Figure 2 concludes with the end of the IGT procedure.

The following sections describe the implementation of the OpenIGTLink module under MeVisLab in detail. The module has been realized as an image processing module (ML) under MeVisLab and the basic source code has been created with the Project Wizard from MeVisLab.

## Implementation of the constructor

The following code example characterizes how the constructor of the OpenIGTLink module has been implemented. The initialization Module(1, 1) in the header of the constructor implementation indicates the number of input and output image connections for the OpenIGTLink module. In our example, we have one input and one output image connection. However, this is not a fixed number, and if a different number is required, the generation of more input and output connections is also possible. ML_TRACE_IN is a status and tracing macro that is set up by default if the Project Wizard of MeVisLab is used for generating the basic source code for a user's own module. The OpenIGTLink module provides several GUI components including an input field for the network port and a button to start listening to the client and setting up the TCP connection. To implement those components, we first suppress calls of the handleNotification function (see next code example) on field changes and to avoid side effects during the initialization phase, we obtain a pointer to the container of all the module's fields. Then, we create different fields, e.g. for the port and the start button. For visualization and further processing of the received data through the OpenIGTLink connection, we additionally create and add an output field for the data of the client (in this case the transformation data). Finally, the calls of the handleNotification function on field changes is reactivated again at the end of the constructor.

```
// constructor
OpenIGTLink::OpenIGTLink()
: Module(1, 1)
{
ML_TRACE_IN("OpenIGTLink::OpenIGTLink()"); // status and tracing macro
// suppress calls of handleNotification on field changes to avoid side
effects during
initialization phase
handleNotificationOff();
// get a pointer to the container of all the module's fields
```

```
FieldContainer *fields = getFieldContainer();
…
// create different fields, e.g. for the port and a start button
_port = fields->addInt("port");
_start = fields->addNotify("start");
// create and add an output field for the transformation data of the client
_transformation = new SoTransform();
(_outSoTransformation =
fields->addSoNode("outputTransformation"))-
>setSoNodeValue(_transformation);
// reactivate calls of handleNotification on field changes
handleNotificationOn();
}
```

## Handling notifications of the OpenIGTLink module

The next code example describes how the important parts of the handleNotification function
for the OpenIGTLink module can be implemented. The handleNotification function is called
on user interactions, e.g. the start button. However, the function is not called when new data
is available through the OpenIGTLink mechanism, that is handled inside the
handleNotification function when the start button has already been pressed. Just as with the
constructor, ML_TRACE_IN is a status and tracing macro that is set up by default if the
Project Wizard from MeVisLab is used for generating the basic source code for a user's own
module. If the start button has been pressed by the user, the initialization for setting up the
connection and the data transfer is prepared. For example, the user-defined port value is
used to create the server socket. Afterwards, if the socket is valid – and therefore the client is
connected – the different data types are checked for incoming data. This procedure follows
the standard way of using the OpenIGTLink library; there are tutorials and a several code
snippets available online (for example, see
http://www.na-mic.org/Wiki/index.php/OpenIGTLink/Library/Tutorial). To keep the code
snippet simple, we only list the *if*-condition for the TRANSFORM data type. If the
TRANSFORM data type has been received from the client, the ReceiveTransform function
is called with the socket und the message header (headerMsg) as parameters.

```
// handle changes of a field
void OpenIGTLink::handleNotification (Field *field)
{
ML_TRACE_IN("OpenIGTLink::handleNotification ()"); // status and tracing
macro
…
if (field == _start)
{
int port = _port->getIntValue();
igtl::ServerSocket::Pointer serverSocket;
serverSocket = igtl::ServerSocket::New();
int r = serverSocket->CreateServer(port);
…
igtl::Socket::Pointer socket;
…
if (socket.IsNotNull()) // if client connected
{
```

```
…
// check data type and receive data body
if (strcmp(headerMsg->GetDeviceType(), "TRANSFORM") == 0)
{
ReceiveTransform(socket, headerMsg);
)
…
}
…
}
…
}
```

## Processing the TRANSFORM data type

The third code example illustrates the implementation of the ReceiveTransform function that is called to handle and process the data type TRANSFORM. The function has two parameters: a socket and the message header that are passed by the function call inside the handleNotification function (see previous section and code example). First, a message buffer is created inside the ReceiveTransform function to receive the transform data. This procedure follows the standard method of using the OpenIGTLink library as is described in the tutorials and complete code examples available online. Next, the notification of the transformation field – that has been created inside the constructor – has to be turned off. Then, the transformation value received by the client is set to the corresponding output field of the OpenIGTLink module. Afterwards, the notification of the transformation field is turned on again. Accordingly, all inventor sensors are forced to be triggered and do a refresh on all viewers. An inventor sensor is an Inventor object that watches for various types of events and invokes a user-supplied callback function when these events occur. Additionally, the transformation field is marked as modified, simulating a change to notify all auditors of the instance.

```
// handle and process data type TRANSFORM
int OpenIGTLink::ReceiveTransform(igtl::Socket * socket, igtl::MessageHeader
*
header)
{
std::cerr << "Receiving TRANSFORM data type." << std::endl;
// create a message buffer to receive transform data
igtl::TransformMessage::Pointer transMsg;
transMsg = igtl::TransformMessage::New();
transMsg->SetMessageHeader(header);
transMsg->AllocatePack();
…
_transformation->enableNotify(false); // turning notification of this field
off
_transformation->setMatrix( SbMatrixValue ); // setting value
_transformation->enableNotify(true); // turning notification of this field on
// force all inventor sensors to be triggered and do a refresh on all viewers
SoDB::getSensorManager()->processDelayQueue(false);
// marks an instance as modified, simulating a change to it this will notify
all
```

```
auditors of the instance

_transformation->touch();

…

}
```

## Description of modules and connections

Figure 3 shows the modules and connections that have been used for realizing the OpenIGTLink protocol under the medical prototyping platform MeVisLab. Overall, the data and communication flow goes from the bottom to the top. It starts with the OpenImage module which can be applied by the user to load an image for example in the DICOM format. After the image is loaded it is automatic passed via an output (right triangle (1)) to the OpenIGTLink module. The main module in the MeVisLab network is the OpenIGTLink module that has several inputs (lower area) and outputs (upper area (2)). One input (the lower left one, triangle (3)) is used for the image data that is provided by an OpenImage module as previously described. The transformation data received from the client is passed on via the third output from the left (upper row) of the OpenIGTLink module to a so called SoGroup module. The transformation data influences a 3D cylinder that is also connected with the SoGroup module. The SoExaminerViewer module in turn gets the transformed cylinder and visualizes it in 3D in a window where several user interactions and settings are possible. An additional transform field is used to analyze the transform data from the client with the decomposed matrices modules. However, these modules are not necessary for the processing of the client data, because it is directly available at one of the module's output. On the right side of the screenshot of Figure 3, the interface of the OpenIGTLink module with all its parameter settings and buttons is shown. The OpenIGTLink module with all its inputs and outputs is shown in detail in Figure 4, where the lower two connections are the inputs. For our implementation, we set up an input for image data and for data structures. The data structures can be used, for example, for maker lists– a marker list is a list of MeVisLab XMarker objects which consists of a 6D Position, a 3D Vector, a Type and a Name property. The upper four connections are the outputs of the OpenIGTLink module. One output, for example, can be used for the image data that has been received by the client via the OpenIGTLink protocol. The next two outputs are OpenInventor outputs that are used to provide only the rotation or the whole transformation for the tracker that has been provided by the client over the OpenIGTLink protocol. Finally, the OpenIGTLink module in the presented example also has an output connection for data structures like marker lists or seed points. Figure 3 shows the interface of the OpenIGTLink module with all its parameters and settings realized as a server, so the tracker clients could connect to it. We also realized our MeVisLab integration as a client. Therefore, the MeVisLab integration connects itself to a sever. We used this integration to connect to a real tracker system from NDI. However, for the realization as a client one more parameter has been used, which was the IP address of the server. For the IP address we set up an additional field in the interface of the OpenIGTLink module where the user can define it manually. After the user starts the client the IP address from the interface field is passed to the handleNotification function (see section *Handling notifications of the OpenIGTLink module*) and is then used to connect the client socket to the server.

## Specification of the main data structures

As previous mentioned, we only list the *if*-condition for the TRANSFORM data type to keep the code snippet simple and clear. However, besides the TRANSFORM data type the OpenIGTLink protocol defines overall five default types: 'IMAGE', 'POSITION', 'STATUS' and 'CAPABILITY'. The main data structures are introduced in the following paragraph, for a detailed description see the publication of Tokuda et al. [20]. The IMAGE format in the OpenIGTLink protocol supports 2D or 3D images with metric information, including image matrix size, voxel size, coordinate system type, position and orientation. The POSITION data type is used to transfer position and orientation information. The data are a combination of three-dimensional (3D) vector for the position and quaternion for the orientation. Although equivalent position and orientation can be described with the TRANSFORM data type, the POSITION data type has the advantage of smaller data size. Therefore it is more suitable for pushing high frame-rate data from tracking devices. The STATUS data type is used to notify the receiver about the current status of the sender. The data consist of status code in a 16-bit unsigned integer, subcode in a 64-bit integer, error name in a 20 byte-length character string, and a status message. The CAPABILITY data type lists the names of message types that the receiver can interpret. Although the OpenIGTLink protocol guarantees that any receiver can at least skip messages with unknown type and continue to interpret the following messages, it is a good idea to get the capability information at system start-up to ensure application-level compatibility of the various devices [20]. All these data structures can be used for the connection between the OpenIGTLink and MeVisLab and therefore transfer different information between a hardware device and MeVisLab based software.

## Results

To use the OpenIGTLink library under the MeVisLab platform (Version 2.1, 2010-07-27 Release), we implemented an ML MeVisLab module in C++ with Microsoft Visual Studio 2008 (Version 9.0.21022.8 RTM). Figure 5 shows a screenshot of the MeVisLab prototype when a client provides tracker coordinates. The tracker coordinates from a tracker client are shown in the command window on the left side of Figure 5. The 3D visualization window in Figure 5 belongs to a SoExaminerViewer of MeVisLab that was used to visualize a cylinder whose location is connected with the tracker coordinates from a tracker client. In our evaluation, we used the simulator programs that come with the OpenIGTLink library. This includes, for example, a TrackerClient that works as a TCP client and is an example that illustrates how to send dummy tracking coordinates to a server. The tracker coordinates from the client could be displayed in real-time with a laptop that has an Intel Atom Z530 CPU, 1.60 GHz, 2 GB RAM, Windows Vista Home Premium x32 Version, Version 2007, Service Pack 1. Note that even if Microsoft Windows Vista is not officially listed under the supported platforms of the OpenIGTLink library, we were able to successfully build, execute and use the simulator programs also under Windows Vista.

The frame rates for the SoExaminerViewer were measured with the SoShowFPS module from MeVisLab. The module was directly connected with the visualization module and superimposes the actual frame per second rate into the rotating tracker. With the introduced laptop configuration, we could achieve 35 fps on the average which is sufficient for the clinical requirements. The detailed times (min, max and mean) in milliseconds (ms) for the transfer and the visualization of 100 packets (via the data type TRANSFORM) from a tracker client to MeVisLab is presented in Table 1 (the overall mean and standard deviation was $30.77 \pm 1.79$ ms). Similar results could be achieved when we send TRANSFORM data types from the laptop to a tracker server – NDI tracking system – over a network. For the transfer and the visualization of 100 packets we measured $19.28 \pm 1.43$ ms (min=17.56 ms and max=24.06 ms). However, this time results also depend on the kind of 3D object that is rendered and visualized – as shown in Figure 5 we visualized a simple cylinder. Therefore, we also measured the time without transferring the data to the visualization module and accomplished $1.48 \pm 0.22$ ms (min=1.35 ms and max=2.86 ms).

## Discussion

In this contribution, we have presented the integration of the OpenIGTLink network protocol for image-guided therapy (IGT) with the medical prototyping platform MeVisLab. MeVisLab is a non open-source framework for the development of image processing algorithms and visualization and interaction methods, with a focus on medical imaging and the OpenIGTLink protocol is a new, open, simple and extensible network communication protocol for image-guided therapy. The protocol allows users to transfer transform, image and status messages as a standardized mechanism to connect software and hardware. To the best of our knowledge, the developed solution is the first approach to successfully integrate OpenIGTLink with MeVisLab. Researchers using MeVisLab now have the possibility to connect to hardware devices that already support the OpenIGTLink protocol, such as the NDI Aurora magnetic tracking system [28]. The integration has been tested with tracker clients that are available online. Another possible application would be the integration of commercial FDA-approved surgical navigation system with research prototype software built on top of the MeVisLab platform [29]. OpenIGTLink communication enables sharing pre- and intraoperative images as well as instrument tracking information between two systems online. The presented integration allows of combination of approved systems and prototype systems that are still under research. This means that researchers can explore new image processing and visualization technique on MeVisLab in the clinical environment, while performing standard surgical planning and image guidance on the commercial system. In fact, an OpenIGTLink interface has become available as an option for research sites in a popular FDA-approved surgical navigation system provided by BrainLAB AG [30].

There are several areas of future work. For example, the latency and CPU load during image data transfers should be evaluated under MeVisLab by varying the data size of images. Moreover, several IGT applications apart from the already available IGT applications, such as the ultrasound navigation system, the tracking devices and navigation software and the MRI-compatible robot system for prostate intervention, will be available soon and therefore should be integrated and evaluated using the OpenIGTLink protocol. Finally, we plan to use the OpenIGTLink protocol for the communication between Slicer and MeVisLab and test the proposed solution under different operating systems such as Linux.

## Acknowledgments

The authors would like to thank Fraunhofer MeVis in Bremen, Germany, for their collaboration and especially Prof. Dr. Horst K. Hahn for his support. This work is supported by NIH 2R01CA111288-06, 5P41RR019703-07, 5P01CA067165-13, 5U54EB005149-07, 5R01CA138586-02, and Center for Integration of Medicine and Innovative Technology (CIMIT) 11-325. Its contents are solely the responsibility of the authors and do not necessarily represent the official views of the NIH.

**Achieved Highlights**

- The successful integration of OpenIGTLink with MeVisLab is presented

- The developed solution allows MeVisLab programs to connect to image-guided therapy (IGT) devices

- Real-time visualization of tracker client information is possible in MeVisLab

- Standardized communication to share target positions, images and device status is provided

- The developed solution enables direct communication between different medical (prototyping) environments, such as Slicer and MeVisLab

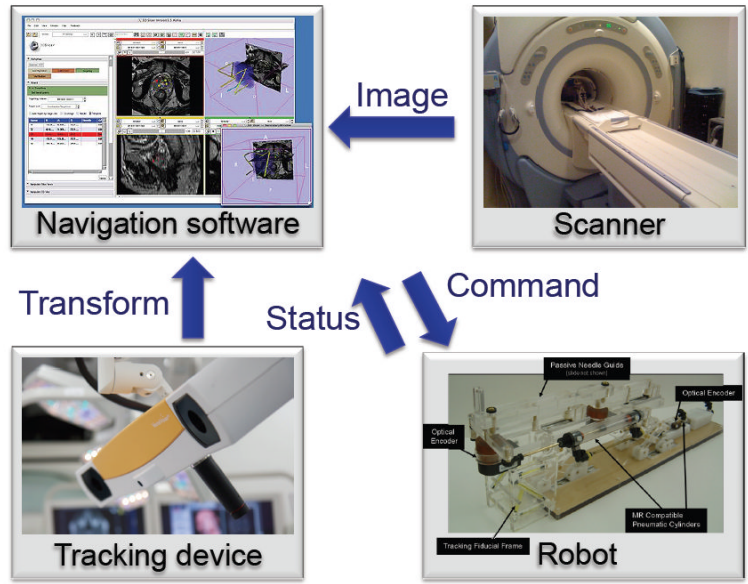

**Figure 1.**
The basic concept of OpenIGTLink (see also
http://www.na-mic.org/Wiki/index.php/OpenIGTLink).

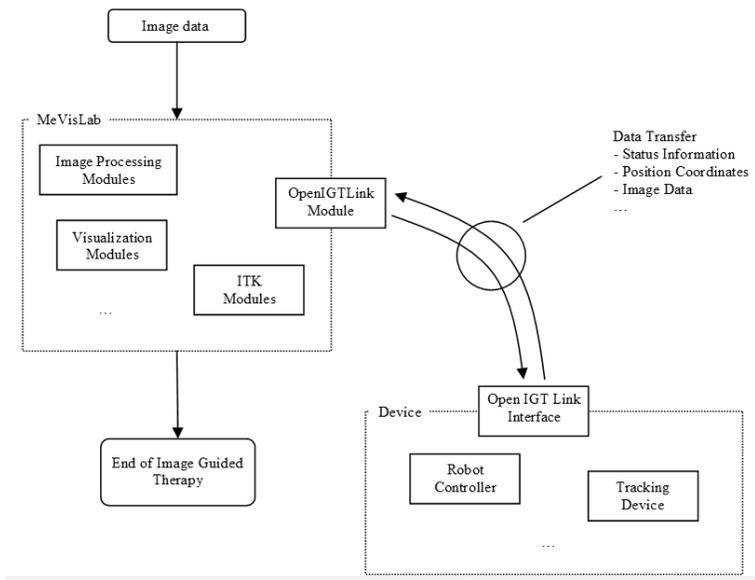

**Figure 2.**
The overall workflow of the presented system, starting with the image data and finishing with the end of Image Guided-Therapy (IGT).

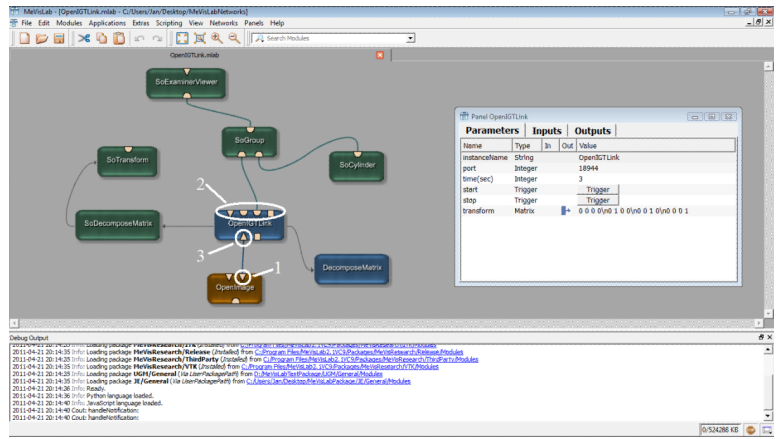

**Figure 3.**
Modules and connections that have been used for realizing OpenIGTLink under the medical
prototyping platform MeVisLab (see http://www.mevislab.de/).

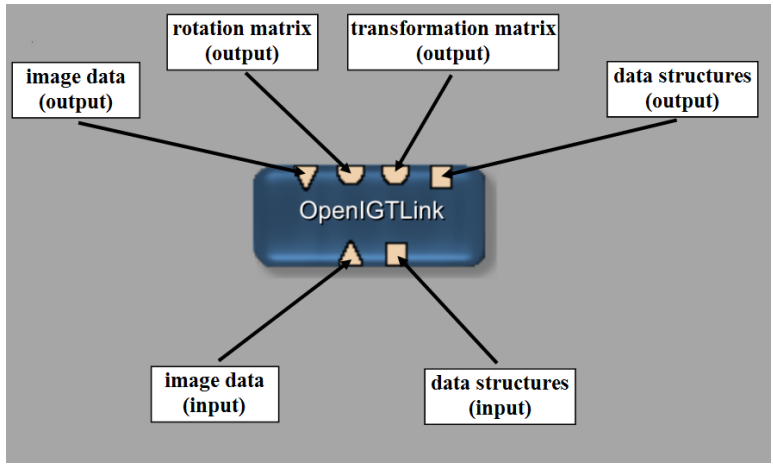

**Figure 4.**
The OpenIGTLink module with its inputs and outputs realized as an ML module in C++ under MeVisLab.

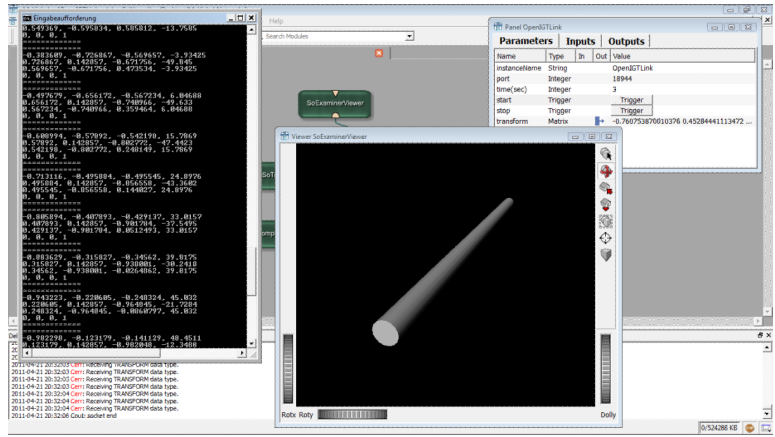

**Figure 5.**
Screenshot of the MeVisLab prototype and a client that provides tracker coordinates.

**Table 1**

Time (min, max and mean) in milliseconds (ms) for the transfer and the visualization of 100 packets (via the data type TRANSFORM) from a tracker client to MeVisLab.

| Run No. | 1 | 2 | 3 | 4 | 5 | 6 | 7 | 8 | 9 | 10 |
|---|---|---|---|---|---|---|---|---|---|---|
| min | 9.34 | 10.76 | 9.39 | 8.15 | 9.71 | 8.39 | 8.25 | 9.69 | 9.46 | 6.66 |
| max | 51.86 | 41.76 | 43.28 | 52.15 | 43.43 | 39.34 | 56.93 | 45.19 | 67.72 | 85.96 |
| mean | 29.66 | 30.14 | 30.01 | 30.24 | 30.09 | 29.98 | 31.86 | 29.70 | 30.45 | 35.54 |